\documentclass[onecolumn,nofootinbib,amsmath,amssymb,amsfonts,superscriptaddress,aps,prfluids,10pt]{revtex4-2}

\usepackage{graphicx}
\usepackage{dcolumn}
\usepackage{siunitx}

\usepackage[executivepaper,margin=0.75in]{geometry}

\usepackage[dvipsnames]{xcolor}

\usepackage{hyperref}
\hypersetup{colorlinks=true,linkcolor=blue,urlcolor=blue,citecolor=blue}

\setcitestyle{authoryear,open={(},close={)}} 
\begin{document}

\title{Solutocapillary bubble centering in a confined ethanol plume in water}

\author{Tobias Baier}
\affiliation{Technische Universität Darmstadt, Fachbereich Maschinenbau, Fachgebiet Nano- und Mikrofluidik, Peter-Grünberg-Straße 10, 64287 Darmstadt, Germany}
\author{Steffen Bisswanger}
\affiliation{Technische Universität Darmstadt, Fachbereich Maschinenbau, Fachgebiet Nano- und Mikrofluidik, Peter-Grünberg-Straße 10, 64287 Darmstadt, Germany}
\author{Sebastian Dehe}
\affiliation{Linac Coherent Light Source SLAC National Accelerator Laboratory, Menlo Park, California 94025, USA}
\author{Steffen Hardt}
\email{hardt@nmf.tu-darmstadt.de}
\affiliation{Technische Universität Darmstadt, Fachbereich Maschinenbau, Fachgebiet Nano- und Mikrofluidik, Peter-Grünberg-Straße 10, 64287 Darmstadt, Germany}




\begin{abstract}
This study investigates the radial centering of gas bubbles within a buoyant plume of ethanol injected into a co-flowing water sheath flow in a vertical capillary. Bubbles nucleate in the ethanol stream due to CO$_2$ supersaturation and rapidly migrate toward the plume axis via solutocapillary (Marangoni) forces driven by interfacial tension gradients in the ethanol-water mixture. Experiments reveal that bubbles of varying sizes reliably align along the plume centerline, facilitated by steep radial concentration gradients near the plume boundary. A reduced-order model supports robust centering across a wide range of bubble radii. For larger bubbles, axial Marangoni effects modulate ascent velocities and can even induce upstream migration under transient conditions, highlighting the complex feedback between bubble dynamics and plume distortion. The results demonstrate that solutocapillary migration provides a reliable mechanism for contact-free bubble focusing, with implications for bubble manipulation in microfluidics, reactors, and phase-separation processes.
\end{abstract}
\maketitle

\section{Introduction}\label{sec:into}

Bubbles are ubiquitous in both nature and technological systems. They can transport mass, momentum and heat over long distances much more efficiently than molecular diffusion \citep{Garbin_2025, Schluter_2021}. They also play a major role in the exchange of gases with the atmosphere in oceans and rivers, where they are generated by braking waves or microbial activity \citep{Emerson_2016}. In both laboratory and industrial contexts, bubbles are frequently created through processes like injection or nucleation. In gas‑liquid reactors or degassing units, bubbles are injected into a liquid medium to introduce gaseous reactants to the liquid phase or to absorb volatile by‑products from the liquid into the gas phase \citep{Kantarci_2005}. During electrolytic reactions, bubbles are generated at the electrode surface and their rapid removal is essential for maintaining high current densities \citep{Park_2023}. In all of these contexts bubbles serve as the means through which a gaseous phase is introduced or removed.

Because bubbles are deformable, buoyant objects, their trajectories are highly sensitive to the surrounding flow field and to interfacial forces \citep{Legendre_2025}. Uncontrolled wall contact can lead to bubble coalescence or attachment, promoting fouling of reactor walls, disturbing the bulk flow, or clogging microfluidic channels, all of which may degrade performance and may cause safety hazards. Consequently, the ability to guide, position, or manipulate bubbles within a flow is a recurring objective in liquid handling or process technology. Strategies that achieve reliable contact-free bubble manipulation are therefore of great practical interest.

The importance of walls and shear stresses in the flow field for the lateral migration of particles in pressure-driven flow along a channel was exemplified by the observation of \citet{Segre_1962} that suspended particles tend to migrate to certain preferred positions at the channel cross section. Being inertial in origin, these lateral migration or lift forces allow sorting and collecting particles in a liquid \citep{DiCarlo_2009}. The analysis of these subtle effects has been extended to wall- and shear-induced migration of bubbles \citep{Legendre_1998, Rivero_2018, Shi_2020}, offering the possibility of manipulating bubbles by tailoring flow fields along a conduit \citep{Hadikhani_2018}. More direct avenues of bubble manipulation are, for example, based on acoustic forces \citep{Baresch_2020}, electric fields \citep{Jones_1977}, or gradients of surface tension \citep{Levich_1969}.

Surface or interfacial tension is strongly dependent on the composition and temperature of the fluid phases in contact. Consequently, any spatial variation of these quantities at the interface creates a tangential stress imbalance that drives Marangoni flow. A drop or bubble dispersed in a liquid of spatially varying composition or temperature therefore experiences an interfacial flow towards larger surface tension, which in turn propels the dispersed phase in the opposite direction. This motion against the surface tension gradient is termed solutocapillary or thermocapillary migration \citep{Subramanian_2002, Lohse_2020}.

A balance between Marangoni stresses and viscous stresses suggests that a characteristic velocity scale for migration of a bubble of radius $R$ placed in a liquid of spatially varying composition is $U_M = -R(d\gamma/d\omega)\nabla\omega/\mu$. Here, $\rho$ and $\mu$ is the density and dynamic viscosity of the liquid, respectively, and the surface tension $\gamma$ is assumed to depend on the mass fraction $\omega$ of one component dissolved in the  liquid. Analytic results are usually obtained assuming constant viscosity and constant $d\gamma/d\omega$, as well as a constant far-field gradient $\nabla\omega$ of the composition. Important parameters influencing the solutocapillary motion are the Reynolds number ${Re=\rho R U_M/\mu}$ and the Marangoni number $\mathit{Ma}=R U_M/D$, identical to the P\'{e}clet number comparing convective and diffusive transport of the spatially inhomogeneous solute with diffusion coefficient $D$. The classical result by \citet{Young_1959}, $U/U_M = 0.5$, for the solutocapillary velocity $U$ of a bubble was obtained in the limit of small $Re$ and small $\mathit{Ma}$, neglecting the influence of the internal gas phase on the outer flow. It was shown by \citet{Balasubramaniam_1987} that this result remains valid for all $Re$ as long as $\mathit{Ma}$ remains small. Extensive simulations for a wide range of values of $Re\leq 2000$ and $\mathit{Ma}\leq 1000$ were conducted by \citet{Balasubramaniam_1989}, indicating that for large $\mathit{Ma}$ the solutocapillary velocity $U/U_M$ saturates at values between 0.15 and 0.25, with a moderate dependence on $Re$. Studies for large $\mathit{Ma}$ by \citet{Crespo_1992} and \citet{Balasubramaniam_1996} have confirmed these asymptotic values analytically. With typical Schmidt numbers $\mathit{Sc}=\mu/(\rho D)=\mathit{Ma}/Re$ of the order of $10^3$ for diffusion in aqueous solutions, the Maragnoni number is much larger than the Reynolds number in such systems. This distinguishes solutocapillary motion from thermocapillary motion in such systems, where the Prandtl number (corresponding to $\mathit{Sc}$) remains below 10.

Here we exploit the variations in surface tension of gas bubbles in an ethanol-water mixture for focusing and guiding them in a core-annular flow of variable composition. In the following section \ref{sec:experiment} we introduce the experimental setup. Bubble centering due to solutocapillary migration is analyzed in section \ref{sec:centering} with a reduced-order model describing the dynamics of this process in section \ref{sec:BubbleCentering_Numeric}. Large bubbles can result in a complex interaction between ethanol plume deformation and Marangoni flow, described in section \ref{sec:plumeDistortion}. We close with a conclusion and outlook in section \ref{sec:conclusions}.

\section{Experimental}\label{sec:experiment}

\begin{figure}[h!t]
	\includegraphics[width=0.8\textwidth]{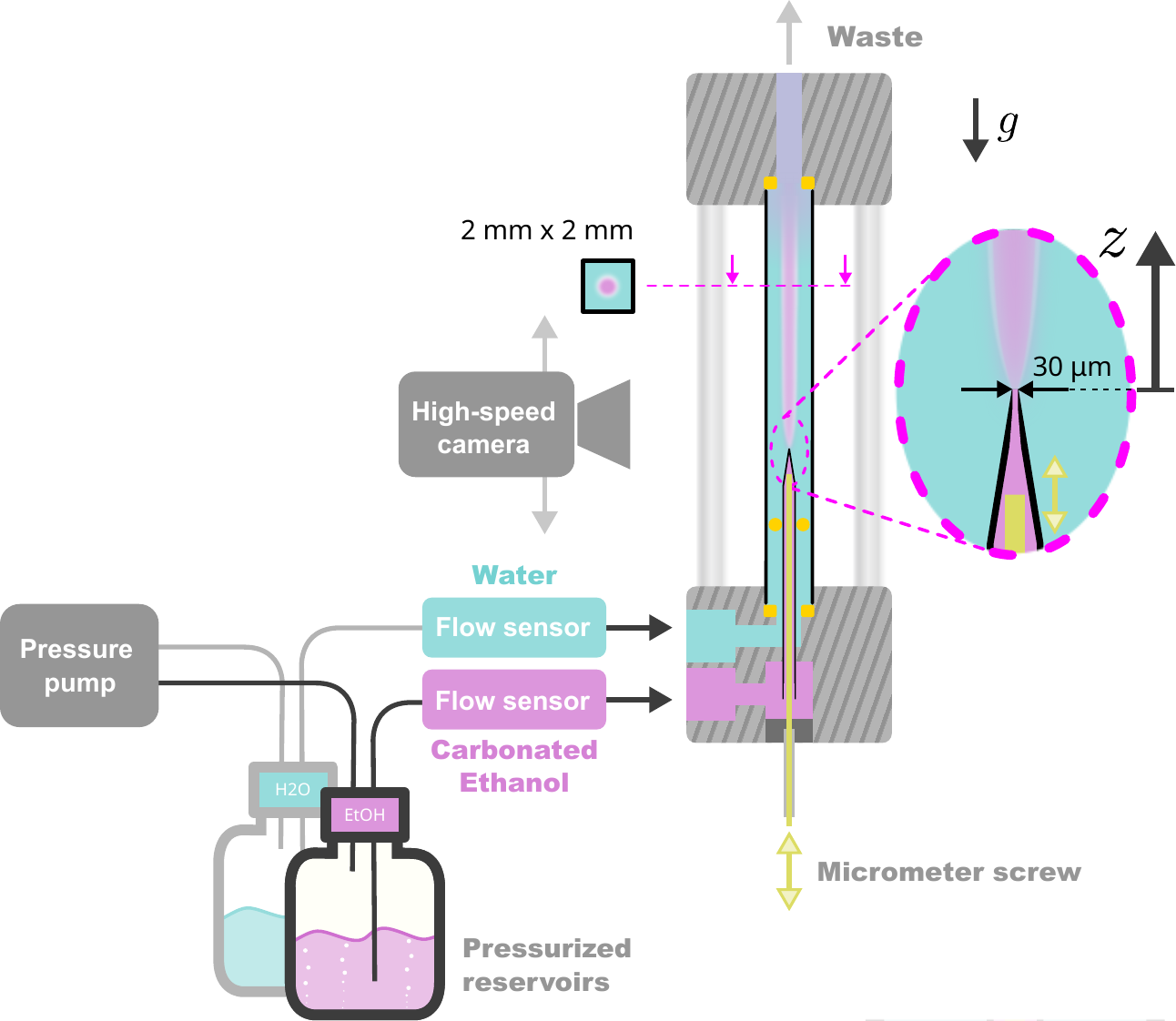}
	\caption{Sketch of the experimental setup. The outer square capillary has inner dimensions of \qty{2}{\milli\metre} by \qty{2}{\milli\metre}. The inner tip diameter of the inner capillary is \qty{30}{\micro\metre}. Ethanol is indicated in lilac and water in blue. The length of the outer capillary is approximately \qty{6}{\centi\metre}.}\label{fig:setup}
\end{figure}

We conduct the experiments in a vertically aligned square‑channel capillary, in which a small inner capillary is arranged coaxially, as sketched in figure \ref{fig:setup}. From the inner capillary a jet of ethanol is injected into a sheath flow of water. At the flow rates chosen in the experiments, ethanol forms a laminar plume taking up a small fraction of the cross section of the outer capillary. As it rises, the plume widens diffusively while remaining confined by the sheath flow at downstream positions where the observations are conducted. Differences in density between the core of the plume and the sheath flow result in a buoyancy-induced deviation of the cross-sectional flow profile from a purely pressure-driven flow, with the ethanol plume ascending within the sheath flow of water. Similar setups were used previously for investigating nucleation, drop formation and drop motion in ternary systems, where mixtures of ethanol with divinylbenzene or transanethole were injected into a sheath flow of water \citep{Hajian_2015, Bisswanger_2026}.

Bubbles are introduced into the ethanol stream by nucleation within the tip of the inner capillary. For this, a narrow constriction introduced into the tip results in a large pressure drop within the ethanol feed flow. Ethanol saturated with CO$_2$ at a higher pressure becomes supersaturated downstream of the constriction, leading to nucleation of CO$_2$ bubbles at the inner wall close to the tip. This allows bubbles smaller than or comparable in size to the opening diameter of the tip to become injected into the ethanol stream, which subsequently grow by further absorption of CO$_2$ from the solution. Varying the degree of supersaturation of the feed flow with CO$_2$ influences the rate and size of bubbles injected into the stream. An example of a chain of bubbles rising within the plume can be seen in figure \ref{fig:radial_marangoni}, which we will turn to after a detailed description of the experimental setup.

\emph{Setup:} The core of the experimental setup is the experimental cell, consisting of an outer capillary with a 2~mm by 2~mm square inner cross-section and an inner capillary with a circular cross-section that converges to an inner diameter of 30~µm at the tip. The outer capillary is clamped between two aluminum blocks using ethylene‑propylene‑diene monomer (EPDM) rubber to provide a chemically resistant seal. The inner capillary is held concentrically inside the outer one using an EPDM rubber ring around the inner capillary. In axial direction it is fixed by a ferrule and nut assembly, commonly used for microfluidic connections. Polytetrafluoroethylene (PTFE) tubing is connected to the aluminum blocks to provide inflow to the cell at the bottom and to route the liquid exiting the cell at the top into a waste container that is ventilated to ambient pressure. To prevent dust particles from entering the inner capillary, a 200~nm in-line PTFE membrane filter is incorporated on the intake side. Flow control for both the outer and inner capillary relies on a pressure pump (Elveflow model OB1 MKII) equipped with a reservoir and microthermal flow sensor for each channel. Prior to filling the reservoir, the liquid can be saturated with CO$_2$ at a pressure greater than 2~bar, which is the limit for the pressure pump. The carbonation is done by injecting compressed gas into the liquid in a sealed bottle, similar to how carbonators for home usage turn tap water into sparkling water. After the carbonation process, the bottle is briefly vented to ambient pressure until connecting it to the pressure pump. The pressure required to achieve a specific flow rate in the inner capillary can be adjusted using a flow resistance built into the inner capillary. For this purpose, a 100~µm thick wire is threaded into the inner capillary. Using a Luer-Lock T-junction, the wire is re-routed away from the liquid intake and to a syringe with a micrometer screw attached, serving as a linear actuator. This allows for precise axial positioning of the wire inside the inner capillary, similar to a miniature version of a Bowden cable system with a needle valve at the end. By closing this needle valve, a large pressure drop (approximately 1~bar as indicated by the pressure pump) is created near the tip in the converging section inside the inner capillary. As a result, the ethanol solution, saturated with CO$_2$, becomes supersaturated, enabling the creation of bubbles small enough to pass through the 30~µm opening. For imaging the flow inside the cell, a high-speed camera (Photron Fastcam Mini AX200) and a long-distance microscope (Navitar 6.5X UltraZoom and Mitutoyo M Plan Apo 10x or 2x) are used. The cell is back-illuminated using an LED panel. To enhance image contrast, a shadow-casting object can be placed between the light source and the experimental cell, effectively creating a similar effect as used in darkfield microscopy. Precise focusing and framing are achieved by translating the experimental cell on a XYZ micrometer stage. The imaging setup allows for recording at length-to-pixel ratios down to approximately 1.4~µm/px and at 6400 frames per second at full image resolution.

\section{Ascend and radial centering of bubbles}\label{sec:centering}

\begin{figure}[h!t]
	\includegraphics[width=0.95\textwidth]{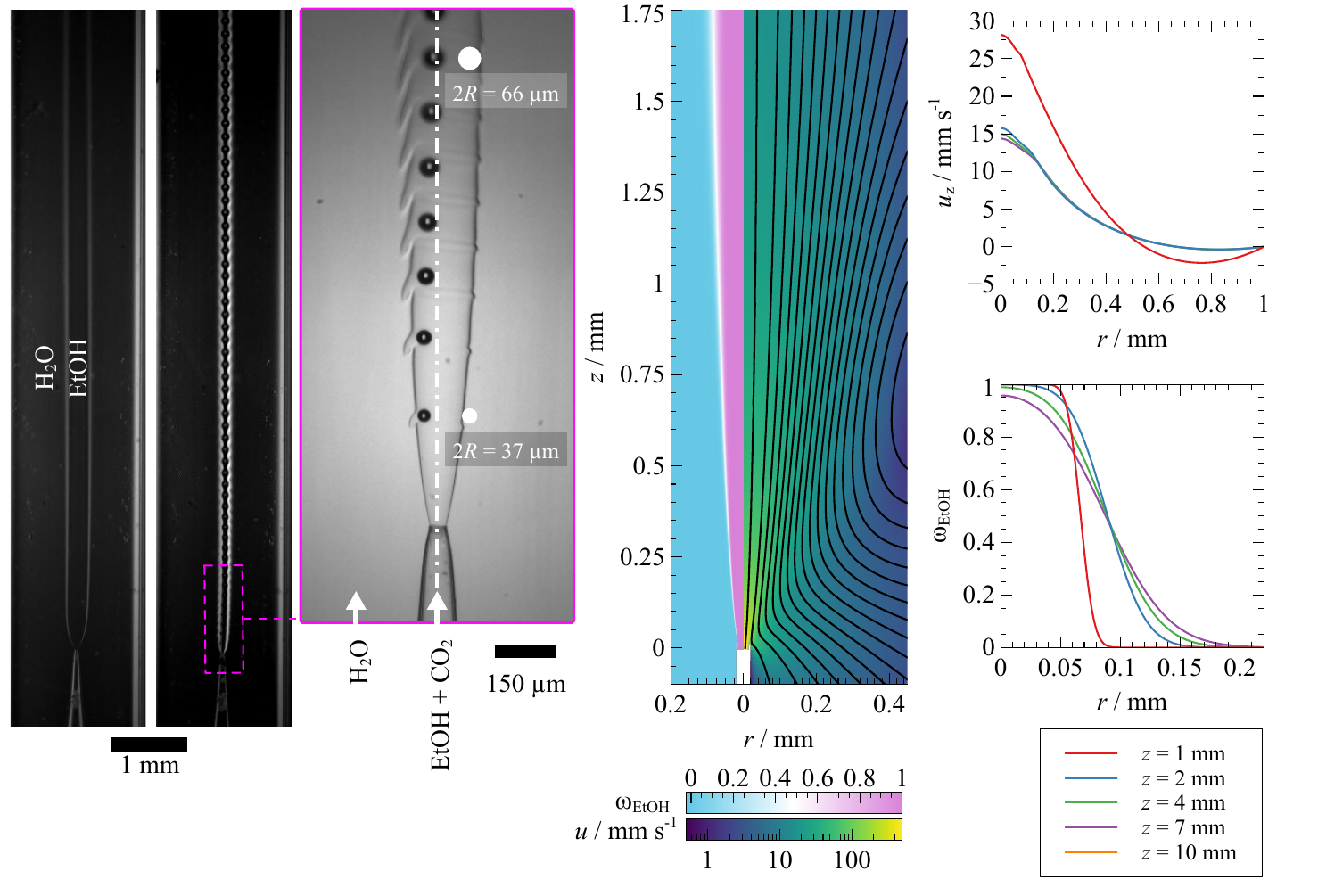}
    \hspace{.5cm}(a)\hspace{1.2cm}(b)\hspace{2cm}(c)\hspace{3.3cm}(d)\hspace{3.2cm}(e)\hspace{1cm}
    \caption{Radial centering of bubbles. (a) pure and (b) carbonated ethanol injected into a sheath flow of water. In (b), a chain of CO$_2$ bubbles rises in the plume of carbonated ethanol. (c) shows a detailed view of the nozzle region indicated by the dashed region in panel (b). Panels (d-e) show simulation results of an ethanol plume injected into a sheath flow of water. The left split in panel (d) shows the ethanol mass fraction, the right split displays the velocity field on a logarithmic scale together with streamlines. In (e), the axial velocity (top) and ethanol mass fraction (bottom) are shown as a function of the radial coordinate, at several positions $z$ downstream of the nozzle. In all cases shown the sheath flow rate was set to $Q_\text{H$_2$O}$ = 200~µl/min, while the ethanol flow rate was $Q_\text{EtOH}$ = 45~µl/min in (a), 42~µl/min in (b), and 23~µl/min in (c-e).}\label{fig:radial_marangoni}
\end{figure}

Figure \ref{fig:radial_marangoni} illustrates some of the flow regimes observed in the experiments. Panels (a) and (b) show a jet of ethanol at flow rates $Q_\text{EtOH} = 45$~µl/min and 42~µl/min, respectively, injected into a sheath flow of $Q_\text{H$_2$O} = 200$~µl/min of water. For the experiment in panel (b), ethanol was saturated with CO\textsubscript{2} at elevated pressure, while in panel (a) ethanol was in equilibrium with air under laboratory conditions prior to the experiment. The difference in optical density of the different fluids allows discerning the edge of the ethanol plume as well as the bubbles under backlight conditions. In both cases, a laminar plume of ethanol rises in the water sheath flow. For the case of carbonated ethanol, a virtually spatially periodic chain of CO$_2$ gas bubbles can be seen ascending together with the ethanol plume. While flow rates are comparable between the two experiments, the width of the plume is different, as the additional buoyancy due to the presence of CO$_2$ bubbles leads to a higher flow velocity at the centerline of the channel, stretching the plume. This interaction becomes less pronounced for lower bubble densities generated at lower supersaturation of the ethanol feed flow.

A striking feature of the rising chain of bubbles is the regular alignment of bubbles along the axis of the plume. To illustrate the release and subsequent radial alignment of bubbles, we show a magnified view around the nozzle region in panel (c), for an experiment with flow rates of $Q_\text{EtOH} = 23$~µl/min and $Q_\text{H2O} = 200$~µl/min; a corresponding video can be found in the supplemental material \citep{SupplementalMaterial}. The CO$_2$ bubbles nucleate at some impurity at the inner wall of the nozzle, usually at the same location and at a very regular rate, and are dislodged and carried along by the ethanol feed while their diameter is smaller than the inner diameter of the nozzle, allowing them to pass the outlet without clogging. Since the bubbles nucleate at the inner wall of the nozzle, they exit at an off-center position right at the edge of the plume. The large concentration gradients between ethanol and water present here lead to a fast diffusional broadening of the plume's edge, such that the bubble quickly experiences a concentration gradient on its surface. As the surface tension strongly depends on the composition of the ethanol-water mixture, with a larger surface tension on the water-rich side, a strong Marangoni flow pushes the bubble towards the center of the plume. At the same time, ethanol-rich fluid is transported radially outward, as evidenced by the protrusion ejected from the plume near each bubble. The protrusion visible on the opposite side of the jet boundary, forming a thin annular band around the jet, is much smaller, since it does not grow due to Marangoni flow but is merely a result of the disturbance of the jet flow as a bubble passes through the nozzle. As the bubble migrates further into the core of the plume, the concentration and surface-tension gradients both become smaller. Nevertheless, even the small gradients in surface tension experienced well inside the plume during the early stages of its diffusional broadening are sufficient to further sustain a radial bubble migration. This is enhanced by the bubble growth in these early stages: absorption of CO$_2$ from the ethanol after ejection from the nozzle leads to an increasing bubble size, which in turn leads to the bubble experiencing an increasing surface tension difference. Under the flow conditions of the experiment, radial Marangoni flow is thus able to center the ascending bubbles along the axis of the plume less than a millimeter away from the nozzle exit.

As our experiments cannot resolve the detailed composition and flow field, we turn to simulations using the finite-element software COMSOL Multiphysics (v6.2) to visualize key qualitative characteristics. Focusing on the unperturbed rising plume, we approximate the outer square capillary by an axisymmetric domain with the same hydraulic diameter, $2R_c$ = 2~mm, and a length of $L_c = 21.5$~mm, aligned with the z-axis. The jet is injected from a concentric capillary with inner and outer diameters of 30~µm and 40~µm, respectively, protruding 1.5~mm into the domain. Volumetric flow rates of $Q_\text{H$_2$O}$ = 200~µl/min and $Q_\text{EtOH}$ = 23~µl/min are prescribed for the sheath flow and injected jet, assuming fully developed flow profiles at the opening of the inner capillary and at the inlet for the sheath flow. A constant pressure is applied at the outlet of the large capillary. We use the Boussinesq approximation \citep{Leal_2007}, representing the ethanol-water mixture as a background fluid of density $\rho_0$, obeying the continuity equation
\begin{align}\label{eq:continuity}
  \mathbf{\nabla} \cdot \mathbf{u} = 0  
\end{align}
and the Navier-Stokes equations
\begin{align}\label{eq:NavierStokes}
  \rho_{0}\left( \frac{\partial}{\partial t}\ \mathbf{u} + \left( \mathbf{u} \cdot \boldsymbol{\nabla} \right)\mathbf{u} \right) = - \boldsymbol{\nabla}p + \boldsymbol{\nabla} \cdot \lbrack\mu(\boldsymbol{\nabla}\mathbf{u} + \left( \boldsymbol{\nabla}\mathbf{u} \right)^{T})\rbrack - \left( \rho - \rho_{0} \right)g\,\mathbf{e}_{z}.  
\end{align}
The concentration-dependent mixture density $\rho$ and viscosity $\mu$ are taken from \citet{Khattab_2012} at temperature $T=298$~K, with the density $\rho_0$ of pure water for the background fluid. In the same vein we approximate the mass transport equation by neglecting the variation in diffusion coefficient and density with composition,
\begin{align}\label{eq:diffusion}
    \frac{\partial}{\partial t}\omega_{\text{EtOH}} + \mathbf{u} \cdot \boldsymbol{\nabla}\omega_{\text{EtOH}} = D \boldsymbol{\nabla}^{2}\omega_{\text{EtOH}},
\end{align}
with $\omega = \omega_{\text{EtOH}}$ being the ethanol mass fraction. We set $D =10^{-9}$~m$^2$/s, a value representing typical binary diffusivities of ethanol and water mixtures in the dilute limit for either component \citep{Parez_2013}. The computational domain was discretized using triangular elements of size 1~µm in the vicinity of the nozzle, gradually increasing to 5~µm in the inner plume region with $r\leq 200$~µm, and with 10~µm on the wall of the outer capillary, while the maximal element size gradually increases up to 50~µm in the rest of the computational domain. Quadratic Lagrange elements were used for the concentration field, as well as for the velocity and pressure fields. Using a segregated approach, the velocity and pressure fields were determined separately from the concentration field in an iterative manner, using the MUMPS direct solver \citep{Amestoy_2001} for each set of discretized equations. Only steady-state solutions are computed.

Results of these simulations are shown in panel (d) of figure \ref{fig:radial_marangoni}, focusing on the region around the nozzle. The ethanol mass fraction is shown on the left, while the right-hand side displays the velocity field together with streamlines. The velocity field is still strongly influenced by the jet inflow within the first 2-3 mm distance from the nozzle. A recirculation region can be seen at larger radial coordinates slightly downstream of the injection nozzle. This is also visible in experiments in the rare occasions when a dust particle or small bubble finds its way into the sheath flow such that its trajectory can be followed.

Further downstream, the flow becomes almost unidirectional, yet it still deviates markedly from a purely pressure‑driven Poiseuille profile because of the mixture's buoyancy, as shown in figure \ref{fig:radial_marangoni}(e). The centerline velocity falls within the 13--16~mm/s\, range, dropping to about 10--11~mm/s towards the edge of the plume, and even reverses into a weak back‑flow beyond about two‑thirds of the channel radius. Immediately downstream of the nozzle, the concentration field shows a rapid radial spread due to convection as the jet velocity decreases. However, at positions beyond $z \gtrsim 2$~mm downstream of the nozzle, the radial concentration profile broadens diffusively, and the initially steep profile gradually evolves into a Gaussian shape, see figure \ref{fig:radial_marangoni}(e) bottom. 

Near the nozzle the concentration field still exhibits a considerably sharper profile. A gas bubble close to the edge of the plume will therefore experience a large variation $\Delta\gamma$ in surface tension. We use values tabulated in \citet{Khattab_2012} to estimate the surface tension of the ethanol-water mixture, neglecting the influence of CO$_2$ on the surface tension. Assuming a radial concentration difference as small as $\Delta\omega_{\text{EtOH}} \sim 10^{-3}$ across a bubble corresponds to a difference in surface tension of $10^{-5}$~Pa~m and a corresponding characteristic radial velocity scale of $\frac{\Delta\gamma}{\mu} \sim 10$~mm/s. A bubble close to the edge of the plume thus experiences a strong Marangoni flow responsible for the radial ejecta of ethanol rich mixture visible in panel (c) of figure \ref{fig:radial_marangoni} next to each bubble. This strong flow ensures that the bubble experiences a sufficiently large radial velocity to reach the center of the plume within the approximately 0.1~s it takes to traverse the field of view in panel (c), with the center of each bubble only traversing a radial distance of less than 50~µm.

\subsection{Reduced-order model: bubble centering}\label{sec:BubbleCentering_Numeric}

In the following, we develop a model for the bubble transport in the plume. The purpose of this model is to capture the essential physics in a description that is as simple as possible. With respect to a comparison with experimental results, we do not aim at a quantitative agreement, but at uncovering some key relationships underlying the bubble dynamics. The corresponding predictions can be tested in future experiments that would usually require a better control of bubble sizes and starting positions than achievable with our setup. 

We neglect the variation of velocity due to buoyancy in the plume and assume that it is transported along the channel at constant velocity $U_0$. Additionally, we neglect the distortion of the concentration field due to the presence of bubbles, in particular due to the Marangoni flow around them. In a co-moving frame of reference with $z$-coordinate $z(t)=U_0 t$, the initial step-like ethanol mass-fraction profile, $\omega_\text{EtOH}(r,t)$, with pure ethanol inside and pure water outside the jet, spreads diffusively as a substance initially confined to a disk on an infinite plane \citep{Crank_1975}, 
\begin{align}\label{eq:diffusion_cylinder}
    \omega_\text{EtOH}(r,t) = \frac{1}{2Dt} e^{-r^2/(4Dt)}
        \int_0^{r_\text{jet}} e^{-s^2/(4Dt)} I_0\left(\frac{rs}{2Dt}\right) s\,ds,
\end{align}
where $I_0(x)$ is the modified Bessel function of the first kind, and $r_\text{jet}$ is the edge of the initially sharp concentration profile.

Classic expressions for the velocity of solutocapillary migration assume a concentration field varying gradually on the scale of a bubble. Particularly in the early stages, our bubbles experience much steeper and inhomogeneous gradients and we therefore turn to an order-of-magnitude estimate of their mobility. The solutocapillary velocity at small Marangoni number \citep{Balasubramaniam_1987}, $U=-R(d\gamma/d\omega)\nabla\omega/(2\mu_0)$, can be interpreted as obtained from a balance between the Hadamard–Rybczynski drag on the bubble in a liquid with viscosity $\mu_0$, $F_D=-4\pi\mu_0 R U$, with the surface-integrated Marangoni stress, $F_M = -\pi R\Delta\gamma$, where $\Delta\gamma=2R(d\gamma/d\omega)\nabla\omega$ is the difference in surface tension over the diameter of the bubble. A characteristic velocity scale for a bubble experiencing a varying surface tension across its diameter thus becomes $U=-\Delta\gamma/(4\mu_0)$, and the radial bubble position in the diffusing plume is thus assumed to obey the kinematic equation
\begin{align}\label{eq:bubbleVelocity}
    \dot{r}(t) = U(r(t)) = -\frac{1}{4\mu_0} 
        \left[\gamma(\omega_\text{EtOH}(r(t)+R,t)-\gamma(\omega_\text{EtOH}(r(t)-R,t)\right],
\end{align}
with $\omega_\text{EtOH}(r,t)$ according to equation \eqref{eq:diffusion_cylinder}.

We integrate equation \eqref{eq:bubbleVelocity} numerically using the LSODA solver \citep{Hindmarsh_1983,Petzold_1983} implemented in SciPy \citep{2020SciPy-NMeth} for different bubble sizes and starting positions. For the surface tension the values reported by \citet{Khattab_2012} at 298~K were taken, while the viscosity and diffusion coefficient were set to $\mu_0 = 10^{-3}$~Pa~s and $D = 10^{-9}$~m$^2$/s. The initial jet radius was set to $r_\text{jet}=100$~µm. For small times, $t<0.01$~s (i.e., $tD/r_\text{jet}^2<10^{-3}$), the concentration profile $\omega_\text{EtOH}(r,t)$, eq. \eqref{eq:diffusion_cylinder}, is approximated by expression \eqref{eq:planarDiffusion} discussed in appendix \ref{sec:shortTimeLimit}.

\begin{figure}[h!t]
	\includegraphics[width=1.\textwidth]{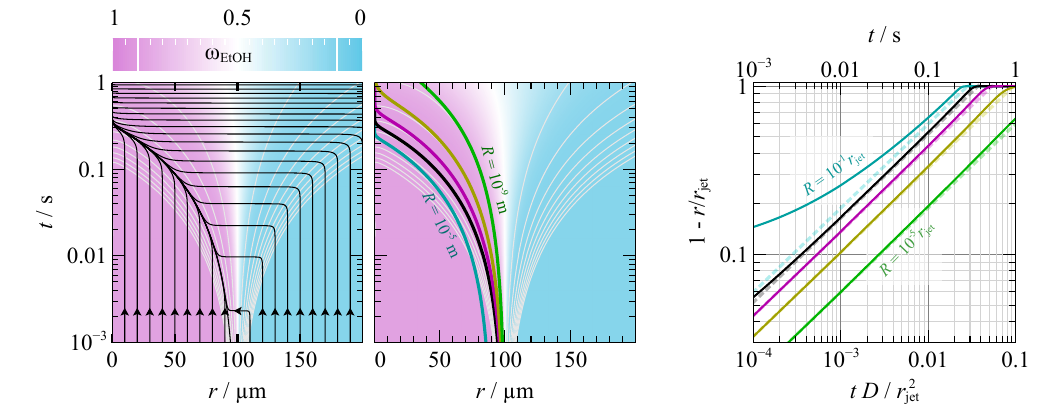}\\
    \hspace{1.cm}(a)\hspace{3.5cm}(b)\hspace{5.cm}(c)
    \caption{Computed radial centering of bubbles in a diffusively broadening jet of initial radius $r_\text{jet}=100$~µm. (a) Trajectories $r(t)$ of bubbles with diameter $R=1$~µm initially located a different radial positions $r_0$ (black lines). (b) Trajectories of bubbles with radii in the range $R=(10^{-9}, 10^{-8}, $\ldots$, 10^{-5})$~m, initially positioned at $r_0 = r_\text{jet}$. The same trajectories are shown in (c) using rescaled nondimensional coordinates. The dashed lines correspond to the short-time limit (\ref{eq:shortTimeSolRadial}--\ref{eq:shortTimeSolRadial_parameters}). In (a) and (b) the ethanol mass fraction $\omega_\text{EtOH}$ is shown in the background with gray isolines at contour levels $10^{-9}, 10^{-8}, \ldots, 10^{-1}$ and $1- 10^{-1},\ldots, 1-10^{-9}$. The contours 0.1 and 0.9 are visible in the colorbar.}\label{fig:centering_numeric}
\end{figure}

Trajectories of bubbles with radius $R=1$~µm, initially released at different starting positions inside and outside the jet, are shown in figure \ref{fig:centering_numeric}(a). Evidently, for very short times only bubbles situated near the sharp initial concentration step experience a significant Marangoni force, and the concentration profile must broaden diffusively before more distant bubbles are affected. Once a bubble experiences a large enough concentration gradient, it is transported towards the center of the jet. This happens slightly differently for bubbles inside the jet than for those outside, but both eventually follow the trajectory of a bubble initially located at the jet radius. Bubbles inside the jet are picked up by the evolving concentration profile, eventually moving along with bubbles released close to the jet radius, while bubbles on the outside are rapidly transported through the diffuse edge of the plume, due to the large concentration and corresponding surface tension differences, until they reach the asymptotic trajectory of a bubble released close to the jet radius. Note that this is even the case for bubbles released at larger distances from the jet, although from a specific point in time on these are rapidly transported all the way towards the center, as they essentially experience a Gaussian concentration profile. Only for very late times and bubbles released far from the center, the concentration gradient may become so weak that a bubble does not reach the channel centerline (not shown). Essentially, our model makes the remarkable prediction that all bubbles released at radial coordinates varying between zero and values significantly larger than the jet radius reach the channel centerline at approximately the same time.

Figure \ref{fig:centering_numeric}(b) shows trajectories of differently sized bubbles, spanning five orders of magnitude from a nanobubble to a 10~µm radius. Since the trajectories of all bubbles released in the vicinity of the jet converge to the trajectory of a bubble released at the initial concentration step, we only plot these asymptotic trajectories here. The rapid centering of the larger bubbles, as observed in experiments, is reproduced in this simple model. However, even nanometer-sized bubbles are attracted to the core of the plume, although it takes substantially longer for them to reach the center. It can be seen that the asymptotic trajectories approximately follow lines of constant concentration. For short times this can be made quantitative, and is explored in Appendix \ref{sec:shortTimeLimit}. In this limit, the asymptotic trajectories are $r(t) = r_\text{jet}-\beta \sqrt{tD}$, with $\beta$ according to \eqref{eq:shortTimeSolRadial_parameters}. This scaling is further investigated in figure \ref{fig:centering_numeric}(c), where we plot $1-r/r_\text{jet}$ as a function of the non-dimensional time coordinate $tD/r_\text{jet}^2$. As can be seen, the short-time approximation excellently predicts the asymptotic curves for the relevant bubble sizes. The deviations observed for large bubbles at short times occur since they are quickly displaced from the initial region of large surface tension gradients, such that they are effectively released with their geometric center at a position $r_\text{jet}-R$. For small bubbles the short-time approximation ceases to be accurate for larger times $tD/r_\text{jet}^2 \gtrsim 0.1$. We have restricted the discussion to bubble radii $R \le 0.1 r_\text{jet}$. Evidently, larger bubbles will significantly influence the concentration field around them. Nevertheless, the concentration gradient will result in a significant inward solutocapillary migration, with rapid bubble centering on short timescales.

\section{Ascending bubble chains and plume distortion}\label{sec:plumeDistortion}

In a typical experiment, the CO$_2$ bubbles originate from a single nucleation site and are produced at nearly constant rate, resulting in an almost spatially periodic chain of bubbles ascending along the center of the plume, as illustrated in figure \ref{fig:chain_velocities} (with corresponding videos in the supplemental material \citep{SupplementalMaterial}). In all cases shown, ethanol is injected at a flow rate of $Q_\text{EtOH}$~=~23~µl/min into a sheath flow of water at a flow rate of $Q_\text{H$_2$O}$~=~200~µl/min. Panels (a) and (c-e) show snapshots of chains of bubbles captured in a window centered at positions $z$~=~2.5--8.5~mm downstream of the nozzle. The bubble size and nucleation rate depends on the nucleation site and the level of supersaturation in the ethanol feed stream, varying with time as the feed saturation decreases due to CO$_2$ lost to the gas phase inside the reservoir. As can be seen by the increase in diameter during ascend, most of the bubbles are still absorbing CO$_2$ from their surroundings at this stage, most prominent for the small bubbles shown in panels (c-e), while the relative change in diameter is imperceptible for the larger bubbles in panel (a).

The dynamics of bubble chains can be visualized by plotting the center pixel column in each frame of a video as a function of time. For the video corresponding to the frame shown in figure \ref{fig:chain_velocities}(a), such a space-time plot is displayed in panel (b). As indicated by the equally spaced parallel dashed lines, the chain of bubbles ascends at a uniform velocity $U_0$ = 15.2 mm/s inside the plume at a center-to-center separation of $l$~=~615~µm. Repeating the analysis for the cases shown in panels (c-e) yields the corresponding velocities and separations indicated on top of each panel. Perhaps surprisingly, despite the variations in bubble size, there is little variation in the velocity of ascend between the different cases. However, this velocity is very similar to the  velocity of an ascending plume of ethanol obtained in the simulations described in the previous section, indicating that convection of the bubbles along with the plume is the dominating transport mechanism, with relative motion of the bubbles within the plume playing a minor role. Nevertheless, particularly for the lager bubbles, their buoyancy, if balanced only by Hadamard–Rybczynski drag, should lead to a velocity of several mm/s relative to the plume, which is not observed. This points to a more intricate interaction between the ascending bubbles and the concentration field around them, with Marangoni flow along the plume's axis reducing a bubble's speed of ascend. Qualitatively, this is explored in a reduced-order model, similar to the one for radial centering, in Appendix \ref{sec:axialBubbleVelocity_reducedOrder}. Diffusion of ethanol from the plume into the surrounding aqueous phase does not only shape the radial concentration profile but also results in an axial concentration gradient with corresponding Marangoni stresses opposing buoyancy. We will come back to this interpretation below.

\begin{figure}[h!t]
	\includegraphics[width=\textwidth]{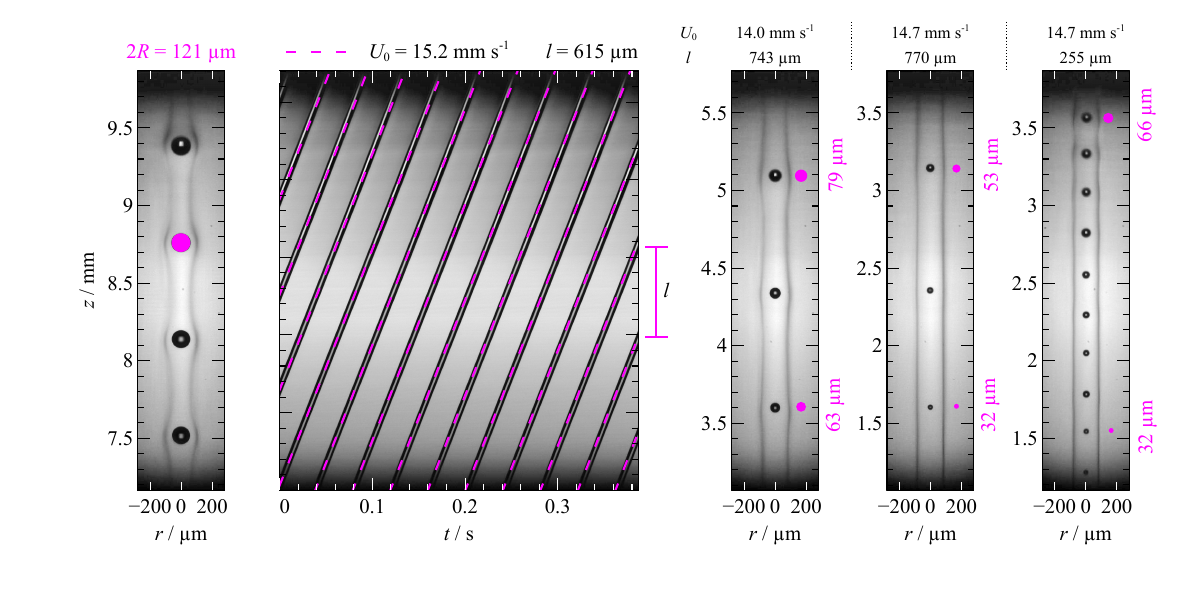}\\
    \hspace{1.5cm}(a)\hspace{3.cm}(b)\hspace{3.5cm}(c)\hspace{1.5cm}(d)\hspace{1.5cm}(e)
	\caption{(a, c--e) CO$_2$ bubbles of various diameters rising inside the ethanol jet. Magenta circles with corresponding label indicate the respective diameters. In (b) a space-time slice of the video corresponding to the frame in (a) is shown. The vertical axis corresponds to the center pixel column in the frame in (a) while the horizontal axis shows time. Parallel straight dashed lines indicate evenly spaced trajectories of a chain of bubbles separated at distance $l$ rising at velocity $U_0$, specified above the figure. For all experiments shown, carbonated ethanol at a flow rate of $Q_\text{EtOH} = 23$~µl/min is injected into a $Q_\text{H2O} = 200$~µl/min sheath flow of water.}\label{fig:chain_velocities}
\end{figure}

As the bubbles increase in size, their influence on the shape of the plume become more pronounced. In particular, the plume is modified by the buoyancy of the bubbles within, as seen by comparing figure \ref{fig:radial_marangoni}(a) and (b), or by bubbles deforming the plume during their ascend, leading to the undulations around each bubble seen in figure \ref{fig:chain_velocities}(a) and (c). Irrespectively, the strong Marangoni centering of the bubbles leads to highly symmetric concentration fields around the axis of the capillary. This is even observed in transient situations with strong modulations of the plume. An example is seen in the time-series shown in figure \ref{fig:axisymmetric_plume} and the accompanying video found in the supplemental material \citep{SupplementalMaterial}, where again carbonated ethanol at a flow rate of $Q_\text{EtOH} = 23$~µl/min is injected into a $Q_\text{H2O} = 200$~µl/min sheath flow of water. Initially, no nucleation site is active, leading to a steady plume of ethanol ascending together with the water. Once a nucleation site becomes active, bubbles start to form. Here the bubbles leaving the nozzle are large enough to momentarily reduce the ethanol flow as they pass through the opening as well as the upstream constriction, in turn modulating the shape of the plume, visible as small undulations originating at the tip. The rapid decrease in velocity as the bubbles exit the nozzle, together with the high nucleation frequency, induces bubble coalescence shortly after injection. The strongly transient nature of these events leads to an irregular plume modulation along its axis, while still maintaining a highly rotationally symmetric concentration field around the axis during the entire process. 

The modulation in plume thickness due to the presence of bubbles affects the concentration that bubbles are exposed to along their meridians (i.e., in flow direction). The corresponding variations in surface tension induce Marangoni flow in meridional direction around the bubble, in turn affecting the plume and resulting in a force on the bubble, this time in axial direction. This flow can be so large that even upstream migration of bubbles can become possible, as shown in the time-series of figure \ref{fig:axisymmetric_plume}. This effect was described earlier in a similar system where instead of bubbles, oil droplets nucleate from a homogeneous ethanol-oil mixture injected into a water stream \citep{Bisswanger_2026}, akin to the nucleation of aniseed oil observed when diluting Ouzo with water. In a similar way as in the Ouzo system, upstream motion is initiated when the jet is deformed into a thin film of ethanol-rich liquid flowing past the bubble. The successive diffusion of water into this thin film leads to a larger surface tension downstream of the bubble, in turn resulting in an upstream migration of the bubble. This inhibits the formation of a quasi-periodic bubble chain and results in growth of the migrating bubble due to coalescence with bubbles emanating from the nozzle. The intricate interplay between Marangoni flow, film thickness and diffusion results in a sustained migration of the bubble down to the nozzle, even as its size and hence buoyancy steadily increases due to bubble coalescence. Note that during the entire process the concentration field and bubble column remain aligned with the axis of the capillary, indicating that Marangoni centering is very robust even under highly transient conditions.

\begin{figure}[ht!]
    \centering
    \includegraphics[width=0.95\textwidth]{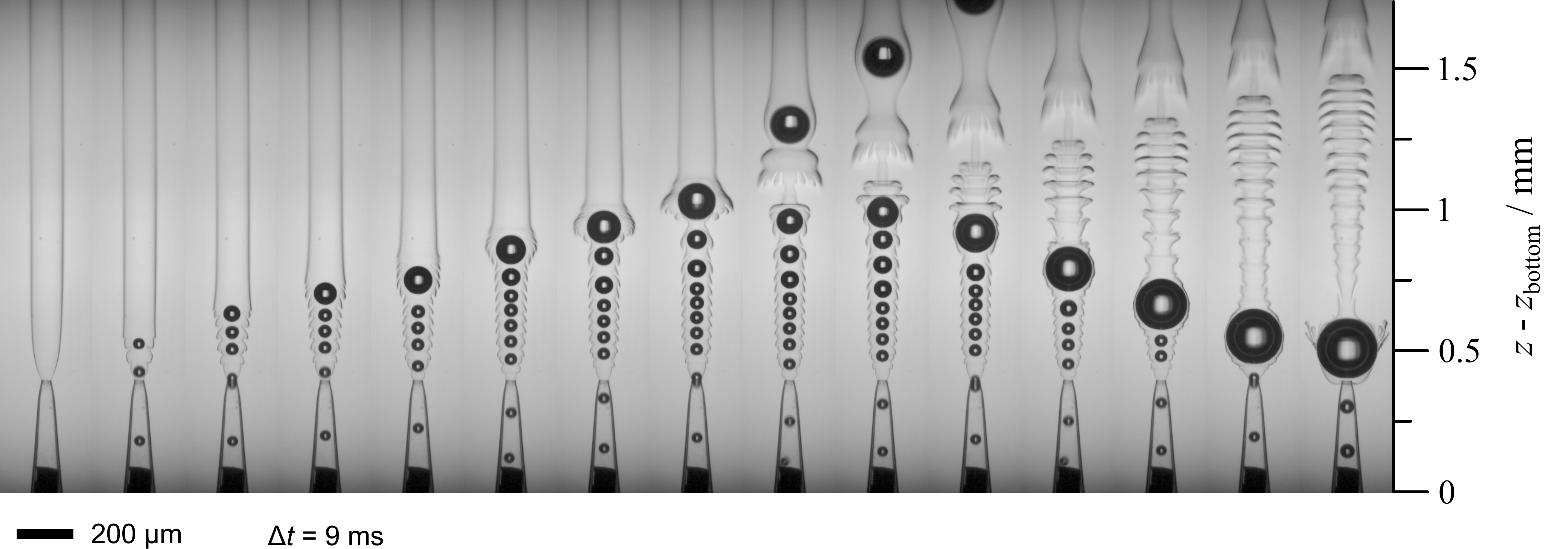}
    \caption{Time-series showing bubble centering and axisymmetric plume modulation under transient conditions. For large enough bubbles, Marangoni flow in axial direction can significantly influence the velocity of ascend, even leading to upstream migration. $Q_\text{EtOH}$ = 23~µl/min, $Q_\text{H$_2$O}$ = 200~µl/min.}\label{fig:axisymmetric_plume}
\end{figure}

We have thus seen that there is a strong interaction between large bubbles and the plume, with bubbles shaping the plume's concentration field, which in turn influences not only the radial motion of bubbles but also their axial motion. While the scenario shown in figure \ref{fig:axisymmetric_plume} is strongly transient, meridonal gradients in surface tension are expected to also affect the velocity of ascend of larger bubbles in the chains shown in figure \ref{fig:chain_velocities}. Particularly larger bubbles are not simply convected along with the plume, but due to their own buoyancy rise with a relative velocity, which suggests that as bubbles grow during their ascend, their velocity increases. Due to the diffusive widening of the plume along the capillary, the upper hemisphere of a bubble experiences a smaller ethanol concentration than its rearward section, inducing an interfacial Marangoni flow in positive $z$-direction along the bubble surface, reducing its rise velocity. This effect increases with bubble size and counters bouyancy-driven transport. While the Marangoni flow cannot be resolved by our experimental techniques, it is plausible this is pivotal for the observed uniform velocity of ascend of the chains with bubbles of different size.

\section{Conclusions and outlook}\label{sec:conclusions}

We have investigated the interaction of bubbles with a plume of ethanol injected into a co-flowing sheath flow of water along a capillary. A striking feature of this interaction is the rapid and robust alignment of bubbles along the axis of the plume, driven by solutocapillary migration in the evolving concentration field. The timescales for bubble centering were investigated using a simple qualitative model, showing that such an alignment is feasible for a large range of bubble sizes. The model predicts that for a given bubble radius the time required for axial alignment is rather independent of the radial starting position of a bubble for a wide range of initial conditions. While the reduced-order model is limited to bubbles small compared to the diameter of the plume, our experiments demonstrate that also much larger bubbles are aligned along the axis of the plume, even though a strong and transient deformation of the plume can ensue due to this interaction. Our experimental setup produces quasi-periodic bubble chains that rise with a uniform velocity. However, the increase in bubble size along their path suggests that the bubble velocity should increase due to buoyancy forces. In that context, it needs to be taken into account that larger bubbles also experience larger Marangoni forces that counter the buoyancy forces. The Marangoni forces can become so strong that a large bubble is transported upstream against the main flow direction.

In our experiments the jet was supersaturated with a dissolved gas, leading to bubble nucleation. In a technical process gas may be produced as a result of a chemical reaction between components present in the liquid. It may also be desirable to inject a gas into a liquid by other means, as such an arrangement promotes gas-liquid reactions or separation by absorption. In such a scenario, a tailored multi-component flow with cross-stream concentration gradients to promote solutocapillary migration could be an effective way to collect and remove bubbles from a stream. As indicated by our numerical results, this is expected to work effectively even when the bubbles are initially present in the sheath flow, as long as the concentration profile becomes wide enough to sufficiently influence the surface tension of bubbles far from the center of the injected jet. Such a scheme can also be useful to avoid bubble contact with walls of a conduit, which may lead to clogging. Moreover, depending on the composition, a scenario with outward migration away from an injected jet can be envisioned, allowing depletion of an unwanted dispersed phase from a product stream. Since surface tension is not only influenced by the liquid composition but also by the temperature profile, an even richer set of operations can be imagined by combining solutocapillary migration with thermocapillary migration.

In this manuscript the focus was on the radial alignment of a dispersed phase along the axis of a jet injected into a sheath flow of a different substance by solutocapillarity migration. However, we have seen that modulations in the plume can impact migration along its axis. It can be envisaged that deliberate shaping of the plume by periodically varying its flow rate may be used to promote bubble coalescence or even intermittent bubble capture similar to the events shown in figure \ref{fig:axisymmetric_plume}.

\begin{acknowledgments}
The authors thank Joachim Groß, Clemens Hansemann, Alexander May and Leon Schuhmann for helping to establish the experimental setup. Financial support by the DFG (Deutsche Forschungsgemeinschaft), Project ID 455566770, is gratefully acknowledged.
\end{acknowledgments}


\appendix

\section{Reduced-order model: short-time limit for bubble centering}\label{sec:shortTimeLimit}

We consider small bubbles initially located at the edge of the ethanol jet. Their velocity, equation \eqref{eq:bubbleVelocity}, can be approximated as (abbreviating $\omega(r,t)\equiv\omega_\text{EtOH}(r,t)$)
\begin{align}\label{eq:bubbleVelocity_gradient}
    \dot{r}(t) = U(r(t)) 
        = -\frac{R\,[d\gamma/d\omega]_{\omega=1}}{2\mu_0} \partial_r \omega(r,t),
\end{align}
where for the change in surface tension with composition, $[d\gamma/d\omega]_{\omega=1}$, it is assumed that the bubble remains on the ethanol-rich side of the diffuse region, in agreement with the numerical results shown in figure \ref{fig:centering_numeric}(b).

For short times, when the diffusion length $\delta = \sqrt{Dt}$ is much smaller than the jet radius $r_\text{jet}$, radial curvature effects become negligible in the diffuse boundary layer between ethanol and water. In this regime, the ethanol mass fraction \eqref{eq:diffusion_cylinder} converges to the classical one-dimensional profile for diffusion across a planar boundary between miscible fluids \citep{Crank_1975},
\begin{align}\label{eq:planarDiffusion}
    \omega(r,t) = \frac{1}{2}\left[1-\text{erf}\left(\frac{r-r_\text{jet}}{\sqrt{4Dt}}\right) \right].
\end{align}
Introducing non-dimensional coordinates $\tilde{t}=tD/r_\text{jet}^2$ and $\tilde{r}(\tilde{t})=r(t)/r_\text{jet}$ together with $\omega(r,t)=\tilde\omega(r/r_\text{jet}, Dt/r_\text{jet}^2)$, equation \eqref{eq:bubbleVelocity_gradient}, becomes
\begin{align}\label{eq:bubbleVelocity_NDim}
    \frac{d}{d\tilde{t}}\tilde{r}(\tilde{t}) = -\frac{R\,[d\gamma/d\omega]_{\omega=1}}{2 \mu_0 D} \partial_{\tilde{r}} \tilde\omega(\tilde{r},\tilde{t})
    = -\frac{\alpha}{\sqrt{4\pi \tilde{t}}}e^{-(\tilde{r}-1)^2/(4\tilde{t})}, 
\end{align}
revealing a characteristic length scale $\Lambda = -2\mu_0 D/[d\gamma/d\omega]_{\omega=1}$, positive for the situation considered. For the ethanol-water system $[d\gamma/d\omega]_{\omega=1}=-16.4$~mPa~m and $\Lambda \approx 1.22\cdot 10^{-10}$~m. In equation \eqref{eq:bubbleVelocity_NDim} we have introduced $\alpha = R/\Lambda$ and inserted $\partial_{\tilde{r}}\tilde\omega(\tilde{r},\tilde{t})$ obtained from \eqref{eq:planarDiffusion}.

\begin{figure}[h!t]
	\includegraphics[width=0.42\textwidth]{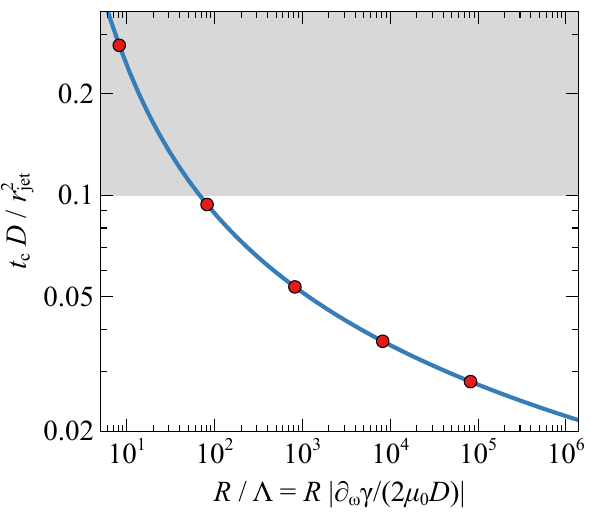}
    \caption{Scaled characteristic time for bubble centering vs.\ scaled bubble radius, as described in the text after equation \eqref{eq:shortTimeSolRadial_parameters}. Circles correspond to the parameter values for the trajectories shown in figure \ref{fig:centering_numeric}(c).}\label{fig:centeringTime}
\end{figure}

The trial solution 
\begin{align}\label{eq:shortTimeSolRadial}
  \tilde y(\tilde t)=1-\tilde r(\tilde t)=\beta\sqrt{\tilde t}
\end{align}
solves \eqref{eq:bubbleVelocity_NDim} when $\beta = (\alpha/\sqrt{\pi})\exp(-\beta^2/4)$. The positive solution of this transcendental equation is 
\begin{align}\label{eq:shortTimeSolRadial_parameters}
  \beta = \sqrt{2W_0(\alpha^2/(2\pi))}, \qquad 
  \alpha = \frac{R}{\Lambda} = -\frac{R [\partial_\omega \gamma|_{\omega=1}]}{2 \mu_0 D},
\end{align}
where $W_0(z)$ is the principal branch of the Lambert W function \citep{Corless_1996}. The corresponding curves are shown as dashed lines in figure \ref{fig:centering_numeric}(c), which agree well with the numerically computed trajectories. Equation \eqref{eq:shortTimeSolRadial} thus is a suitable approximation for the trajectories of a bubble initially located near the edge of the jet, as long as the bubbles are small enough, $R\ll r_\text{jet}$, and for sufficiently small times $t\ll  r_\text{jet}^2/D$. Note that here we do not assume that the bubble is small compared to the width of the diffusion zone. As a consequence, the nonlinearities of the concentration profile are not captured here. In the numerically evaluated reduced order model in section \ref{sec:BubbleCentering_Numeric}, this was incorporated by taking a finite difference scheme. Based on these approximations, a characteristic timescale for centering of such a bubble is $\tilde{t}_c = Dt_c/r_\text{jet} = \beta^{-2}$. This timescale is plotted in figure \ref{fig:centeringTime} as a function of $R/\Lambda$. Since the approximation \eqref{eq:planarDiffusion} for the concentration field \eqref{eq:diffusion_cylinder} becomes increasingly imprecise beyond $\tilde{t} \gtrsim 0.1$, the shaded region in this plot indicates where the estimated $\tilde t_c$ becomes progressively unreliable.

\section{Reduced-order model: axial velocity}\label{sec:axialBubbleVelocity_reducedOrder}

The influence of Marangoni stresses and buoyancy on the axial velocity of a centered bubble relative to the velocity of the buoyant plume can be estimated in a similar manner as the radial velocity. An analytic estimate for the conentration gradient in axial direction can be obtained by noting that on the centerline of the jet, equation \eqref{eq:diffusion_cylinder} for the mass fraction $\omega_\text{EtOH}(0,t)$ simplifies to \citep{Crank_1975} 
\begin{align}\label{eq:diffusion_cylinder_centerline}
    \omega_\text{EtOH}(0,t) = 1 - e^{-r_\text{jet}^2/(4Dt)}.
\end{align}
Correspondingly, since the concentration gradient in axial direction is much smaller than in radial direction, the force due to Marangoni stress can then be approximated as
\begin{align}
    F_M \approx - 2\pi R^2 \left.\frac{d\gamma\hfill}{d\omega_\text{EtOH}}\right|_{\omega_\text{EtOH}=1} \frac{d}{dz}\omega_\text{EtOH}(0, z/U_0),
\end{align}
where the derivative of the surface tension $[d\gamma/d\omega_\text{EtOH}]_{\omega_\text{EtOH}=1}=-16.4$~mPa~m is evaluated for large ethanol mass fractions on the jet axis. Balancing $F_M$ with the Hadamard–Rybczynski drag $F_D$ and the buoyancy force $F_B = \frac{4\pi}{3}R^3 g\Delta\rho$, we obtain for the bubble velocity relative to the characteristic velocity $U_0 \approx 15$~mm/s of the plume
\begin{align}\label{eq:bubbleVelocity_axial}
    U_\text{rel} = \frac{\Delta\rho g R^2}{3\mu_0}\left(1-\frac{3 [d\gamma/d\omega_\text{EtOH}]_{\omega_\text{EtOH}=1}}{2\Delta\rho g R} \frac{d}{dz}\omega_\text{EtOH}(0, z/U_0) \right).
\end{align}

\begin{figure}[h!t] 
    \includegraphics[width=0.5\textwidth]{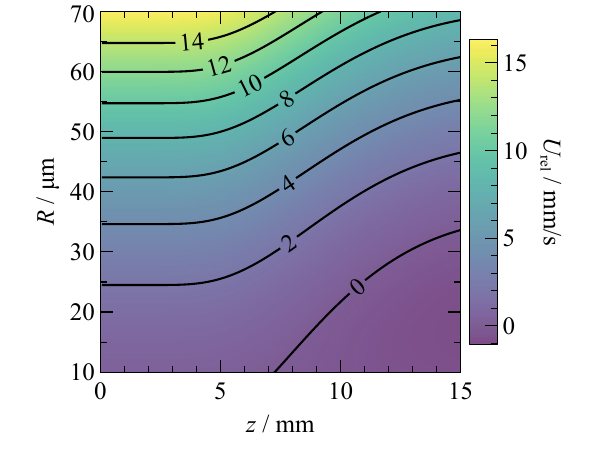}
    \caption{Relative velocity $U_\text{rel}$ with respect to the characteristic jet velocity $U_0=15$~mm/s due to a balance of drag, Marangoni and buoyancy forces according to equation \eqref{eq:bubbleVelocity_axial}.}\label{fig:axial_reduced}
\end{figure}

A contour plot of $U_\text{rel}(R,z)$ as a function of bubble radius $R$ and axial position $z$ is shown in figure \ref{fig:axial_reduced}, where the product of acceleration due to gravity $g$ and density difference between liquid and gas phase, $\Delta\rho$, was set to $g\Delta\rho = 10^4$~N/m$^{3}$. Since we have focused on the axial gradient in the center of the plume to estimate the Marangoni stresses, the model overpredicts the relative ascend velocity for short times (small $z$), as it takes some time for the evolving profile to result in a significant gradient there. Particularly larger bubbles will experience the evolving gradient earlier, such that the axial Marangoni stresses become significant at shorter times (smaller $z$). Additionally, larger bubbles influence the shape of the plume and experience the variation in velocity across the plume's cross section, slowing them down further with respect to the plume's centerline velocity. Despite these shortcomings, the model qualitatively captures the fact that axial Marangoni forces can markedly influence bubble ascend. This supports the hypothesis that a delicate balance between buoyancy, drag, and the concentration field surrounding the bubbles, producing Marangoni flow, is responsible for the modest velocity differences observed in figures \ref{fig:chain_velocities}(a-e). Specifically, figure \ref{fig:axial_reduced} indicates that there are trajectories in the $(R,z)$-space for which $U_\text{rel}$ is independent of $z$, which means that the bubble size increases at a rate at which the changes in buoyancy and Marangoni forces exactly cancel. However, a direct comparison with experimental observations is hampered by the drastic simplifications inherent in the model and is further exacerbated by the fact that the buoyant bubble train influences the characteristic plume velocity, which is difficult to measure independently in our experiments.

\bibliography{literature}

\end{document}